\documentclass[
    prl, twocolumn, superscriptaddress, nofootinbib, amsmath, amssymb,
    aps, floatfix
]{revtex4-2}
\usepackage{graphicx}
\usepackage{latexsym}
\usepackage{slashed}
\usepackage{bm}
\usepackage{color}
\usepackage[colorlinks,citecolor=blue,urlcolor=blue,linkcolor=blue]{hyperref}
\usepackage{diagbox}
\def\({\left(}
\def\){\right)}

\def\beq{\begin{equation}}
\def\eeq{\end{equation}}

\begin{document}

\title{Gravitational Wave as a Probe of Light Feebly Interacting Dark Matter}

\author{Yuchao Gu}
%\email{guyc@pmo.ac.cn}
\email{guyc@njnu.edu.cn}
\affiliation{Department of Physics and Institute of Theoretical Physics, Nanjing Normal University, Nanjing, 210023, China}
\affiliation{Key Laboratory of Dark Matter and Space Astronomy, Purple Mountain Observatory, Chinese Academy of Sciences, Nanjing 210023, China}

\author{Liangliang Su}
\email{liangliangsu@njnu.edu.cn}
\affiliation{Department of Physics and Institute of Theoretical Physics, Nanjing Normal University, Nanjing, 210023, China}

\author{Lei Wu}
\email{leiwu@njnu.edu.cn}
\affiliation{Department of Physics and Institute of Theoretical Physics, Nanjing Normal University, Nanjing, 210023, China}

\author{Yongcheng Wu}
\email{ycwu@njnu.edu.cn}
\affiliation{Department of Physics and Institute of Theoretical Physics, Nanjing Normal University, Nanjing, 210023, China}

\author{Bin Zhu}
\email{zhubin@mail.nankai.edu.cn}
\affiliation{Department of Physics, Yantai University, Yantai 264005, China}

%\date{\today}% It is always \today, today,
             %  but any date may be explicitly specified

\begin{abstract}

Light feebly interacting dark matter is widely predicted in a plethora of new physics models. However, due to very feeble couplings with the Standard Model particles, its relic density produced via the relativistic thermal freeze-out process easily exceeds the observed value. The entropy dilution in an early matter-dominated era provides an attractive mechanism for solving such an overabundance problem. In this work, we note that this dark matter dilution mechanism will lead to two distinctive kinks in the primordial GW spectrum, whose frequencies strongly correlate with the DM mass. We show that the GW detectors, such as Cosmic Explorer (CE) and Big Bang Observer (BBO), can measure the kinks in the primordial GW spectrum and will offer a new avenue to probe light feebly interacting dark matter.

\end{abstract}
%\pacs{Valid PACS appear here}% PACS, the Physics and Astronomy
                             % Classification Scheme.
%\keywords{Suggested keywords}%Use showkeys class option if keyword
                              %display desired
\maketitle

%\tableofcontents

\section{Introduction}
The Weakly Interacting Massive Particle (WIMP) has long been considered a compelling candidate for DM and has been extensively searched for in collider, direct, and indirect detection experiments~\cite{XENON:2019rxp, CDEX:2019hzn, PandaX-II:2016wea, LUX-ZEPLIN:2022xrq, LUX:2016ggv, Super-Kamiokande:2015xms, Fermi-LAT:2014ryh, Cui:2016ppb, CMS:2018ffd}. However, despite these efforts, no unambiguous signal confirming the existence of WIMP has been observed. Conversely, these experimental observations have imposed stringent limits on the parameter space of WIMP~\cite{Arcadi:2017kky, Aprile:2018dbl}. Consequently, light feebly interacting DM candidates, such as the gravitino and axino~\cite{Moroi:1993mb, Covi:1999ty, Covi:2001nw, Roszkowski:2004jd, Bae:2014efa}, have gained significant attention due to their ability to evade the constraints from direct detection experiments, where their super-weak coupling to visible particles makes them unlikely to be probed through conventional direct detection searches. However, the light feebly interacting DM can be probed at the colliders~\cite{Co:2015pka,Brandenburg:2005he}, but also through using telescopes and neutrino experiments to detect the decay products of parent particle~\cite{Choi:2010jt,Gomez-Vargas:2019vci,Gomez-Vargas:2019mqk,Gu:2021lni}.  Furthermore, the primordial GW can be regarded as a probe that detects high-energy physics from the end of inflation to Big Bang Nucleosynthesis. The primordial GW therefore can be utilized to probe such light feebly interacting DM.

During the early high-temperature phase of the Universe, light feebly interacting DM would have been brought into thermal equilibrium with the surrounding particles. As a result, the relic density of such DM, produced through thermal production mechanisms~\cite{Bolz:2000fu, Pradler:2006qh, Pradler:2006hh, Co:2016fln, Co:2017orl} could potentially lead to an overclosure of the Universe due to its small annihilation rate compared to WIMP DM~\cite{Asaka:2000zh, Cheung:2011mg}. Intriguingly, the overproduced abundance of DM can be diluted by the additional production of entropy, aligning it with the observed value. This additional entropy production can be generated by the late decays of certain states~\cite{Moroi:1999zb, Thomas:1995ze, Moroi:1994rs, Allahverdi:2002nb, Moroi:2002rd, Allahverdi:2002nb, Lahanas:2011tk, Fujii:2002kr,Cosme:2020mck, Co:2015pka, Gu:2020ozv, Gu:2021lni, Nemevsek:2022anh}, which often give rise to
an early matter-dominated era (EMD),
preceding the regular radiation-dominated epoch.

Furthermore, the recent milestone discovery of gravitational waves (GWs) has opened up a new era in astronomy and cosmology~\cite{LIGOScientific:2014pky, VIRGO:2014yos, LIGOScientific:2016aoc}. GWs induced by cosmic strings~\cite{Nielsen:1973cs, Kibble:1976sj, Hindmarsh:1994re}, primordial inflation~\cite{Abbott:1984fp} and strong first-order phase transition~\cite{Weir:2017wfa,Athron:2023xlk} offer novel cosmological probes that provide an exciting opportunity to investigate the early Universe's history~\cite{Turner:1993vb, Boyle:2005se, Watanabe:2006qe, Kuroyanagi:2011fy, Kuroyanagi:2018csn, Bernal:2019lpc, Boyle:2007zx,  Cui:2017ufi, Caprini:2018mtu, Cui:2018rwi, Bertone:2019irm, Yuan:2021ebu, Tsukada:2020lgt, Chatrchyan:2020pzh, Samanta:2021mdm, Borah:2022byb, Cui:2017ufi, Cui:2018rwi, Gouttenoire:2019kij, DEramo:2019tit, Borah:2022byb}, including an EMD era. The EMD era can modify the GW spectrum, thus offering the potential to search for light feebly interacting DM through the detection of GW signals.

In this paper, we investigate the role of inflationary GWs as a probe for light feebly interacting DM. Notably, the EMD era induces distinctive features in the primordial GW spectrum where there exist two kinks. The frequencies of these kinks are determined by parameters associated with the EMD epoch. Additionally, these parameters govern the magnitude of the dilution effect occurring during the EMD era and are connected to the mass of the DM candidate through the overproduced DM abundance. Consequently, the EMD era serves as a bridge between GWs and DM, providing a unique opportunity to probe light feebly interacting DM across different mass ranges using inflationary GW observations. By exploring the relationship between the EMD era, the primordial GW spectrum, and the properties of light feebly interacting DM, we aim to uncover new insights into the nature of DM and its connection to the early Universe. Our study highlights the potential of inflationary GW observations as a powerful tool for investigating the properties of DM, offering a complementary approach to traditional collider and (in)direct detection experiments.

\section{The DM relic density and Early Matter-Dominated Era}
\label{section2}
In this work, we consider that the reheating temperature after inflation $T^{I}_R$ is larger than the freeze-out temperature of light feebly interacting DM $T_f$. Therefore, despite its feeble interactions with standard model particles, light feebly interacting DM would freeze out in the early Universe due to the compensation of high reheating temperature $T^{I}_R$. However, due to its feeble interaction, light feebly interacting DM decouples from the thermal bath in an early time when the freeze-out temperature of light feebly interacting DM is still much higher than its rest mass which provides the DM relic density as
% results in the observed DM relic density,
\begin{equation}
    \Omega_{\chi} h^{2}= \frac{m_\chi Y_{\infty} s_{0}}{\rho_c} h^2 \approx 0.12 \left(\frac{m_\chi}{112\, \rm{eV}}\right),
\end{equation}
Here, $\Omega_{\chi} h^{2}$ represents the present-day DM relic density, $m_\chi$ is the DM mass, $Y_{\infty}$ is the present-day DM yield, $s_0$ is the current entropy, $\rho_c$ is the critical density, and $h$ is the reduced Hubble constant. The DM yield is given by
\begin{equation}
    Y_{\infty}=\frac{135 \zeta(3)}{8 \pi^4} \frac{g_\chi}{g_s(T_f)},
\end{equation}
where $g_\chi$ represents the internal degrees of freedom (d.o.f) of DM, $g_s(T)$ is the effective number of relativistic d.o.f of entropy at temperature $T$, and $T_f$ is the freeze-out temperature of DM. In contrast to WIMP DM, the relic density of light feebly interacting DM can easily exceed the observed value of $\Omega_{\chi}^{\rm{obs}} h^{2}=0.12$ due to its weaker annihilation cross-section. To reconcile this discrepancy, the overproduced DM abundance can be diluted by extra entropy production, which is usually induced by a late-decaying state, resulting in an EMD era. It is important to note that the EMD scenario is merely one among various possibilities for explaining the observed relic density of dark matter. Additional mechanisms, including late-time entropy production from alternative sources or non-standard interactions of dark matter, can also be taken into account. However, the key aspect that entropy dilution is correlated with the mass of dark matter remains independent of specific model details.

In this study, we consider the moduli as the late-decaying state~\cite{Evans:2019jcs}. The evolution of the early Universe during the EMD era is governed by the coupled Boltzmann equations,
\begin{equation}
\begin{aligned}
    \frac{d\phi_{m}}{dt}&=-\Gamma_{m} \phi_m,\\
    \frac{d\phi_{R}}{dt}&=a \Gamma_{m} \phi_m,
\end{aligned}
    \end{equation}
where $\phi_{R}$ and $\phi_{m}$ represent the co-moving energy density of radiation and moduli, respectively. $\Gamma_m$ is the decay width of moduli, $a$ is the scale factor, and $M_{\rm{Pl}}$ is the reduced Planck mass. The Hubble expansion rate during the EMD era is given by
\begin{equation}
    H^2=\frac{1}{3 M_{\rm{Pl}}^2}\left(\frac{\phi_{R}}{a^4}+\frac{\phi_{m}}{a^3}\right),
\end{equation}

The EMD era consists of adiabatic and non-adiabatic phases~\cite{Co:2016fln,Co:2017orl}, starting at the cosmic temperature $T_{\rm{eq}}$ and ending at $T_{R}$. These characteristic temperatures, determined by the decay width and co-moving energy density of the moduli, can be expressed as follows
\begin{equation}
    T_{\rm{eq}}=\left(\frac{30 \phi_m}{\pi^2 g_{\rho}(T_{\rm{eq}})}\right)^{1/4},
    \label{Teq}
\end{equation}
\begin{equation}
    T_{R}=\left(\frac{90}{8 \pi^3 g_{\rho}(T_R)}\right)^{1/4} \sqrt{\Gamma_m M_{\rm{Pl}}}.
    \label{TR}
\end{equation}
Here, $g_{\rho}(T)$ represents the effective number of relativistic d.o.f of the energy density at the cosmic temperature $T$. The moduli decay injects additional entropy into the early Universe, and the resulting dilution effect can be parameterized by the dilution factor
\begin{equation}
    D \equiv \frac{S_{\rm{after}}}{S_{\rm{before}}}=\frac{g_s(T_R) T_R^3}{g_s(T_{\rm{eq}})T_{\rm{eq}}^3} \frac{a(T_R)^3}{a(T_{\rm{eq}})^3}=\frac{g_s(T_R) g_{\rho}(T_{\rm{eq}})}{g_s(T_{\rm{eq}}) g_{\rho}(T_R)} \frac{T_{\rm{eq}}}{T_R},
    \label{dilution}
\end{equation}
where $S_{\rm{before}}$($S_{\rm{after}}$) is the co-moving entropy of the Universe before(after) moduli decay. We use the relation between the scalar factor and Hubble expansion rate in EMD era $a(T) \propto H(T)^{-2/3}$ to derive the dilution factor.  It is worth noting that the dilution factor is valid for DM production before the beginning of the non-adiabatic EMD era. As the freeze-out temperature of light feebly interacting DM is very high, the dilution from Eq.~\ref{dilution} is always valid in our scenario. The thermally overproduced DM abundance with different DM masses $m_\chi$ needs the suitable dilution factors $D=m_\chi/112$ to reconcile with the observations, where $m_\chi$ is in the unit of eV. For example, the Lyman-$\alpha$ forest data sets put strong constraints on the free streaming of warm dark matter, further on the mass of a thermal relic warm dark matter $m_{\chi} > 5.3$ keV~\cite{Irsic:2017ixq}, which implies that the dilution factor $D \approx 47$ at least is required to satisfy the warm dark matter thermal relic.
The larger the DM mass $m_\chi$ is, the larger the required dilution factor is, as the thermally produced DM yield is proportional to $m_\chi$. Besides, the entropy injection can not only dilute the DM thermal relic but also cool down the DM velocity. The DM particle with mass $m_{\chi}$ that is thermal freeze-out when relativistic has a present-day velocity,
\begin{equation}
    \langle v_{\chi}^0 \rangle \approx 0.023 {\rm{kms^{-1}}} \left(\frac{g_{\rho}(T_f)}{100}\right)^{-1/3}\left(\frac{m}{1{\rm{keV}}}\right)^{-1}.
\end{equation}
The limits between hot, warm, and cold dark matter are ambiguous. One can qualify as warm dark matter with velocity $0.0018$ kms$^{-1} \leq v_{\chi}^{0} \leq$ 0.054 kms$^{-1}$ ~\cite{Jedamzik:2005ir}. The extra entropy injection after the DM thermal freeze-out cools down the DM velocity:$\langle v_{\chi}^0 \rangle \rightarrow \langle v_{\chi}^0 \rangle/D^{1/3}$. This indicates that the dilution factor $D \approx 72$ is required to satisfy warm dark matter bounds when $m_\chi=0.1$ keV. Note that the dilution factor $D$ relies on these two characteristic temperatures $T_R$ and $T_{\rm{eq}}$, which are related to the EMD era induced by the moduli. Intriguingly, the EMD era will cause two turning points in the primordial GW spectrum, whose frequencies are also dependent on these two characteristic temperatures. This indicates that the primordial GW spectrum can be utilized to look for light feebly interacting DM. In the next section, we will evaluate the primordial GW spectrum during an EMD era and discuss the prospect of DM detection through primordial GWs.

\section{The Primordial Gravitational Waves Spectrum in EMD Era}
\label{section3}
Primordial gravitational waves represent tensor perturbations in a spatially-flat Friedmann-Robertson-Walker (FRW) universe, described by the metric,

\begin{equation}
    ds^2=-dt^2+a^2(t)(\delta_{ij}+h_{ij}(t,\Vec{x})) dx^i dx^j,
\end{equation}
Here, the gauge-invariant tensor perturbation $h_{ij}$ satisfies $h_{ij}=h_{ji}$, along with the transverse and traceless conditions: $h_{ii}=0$ and $h_{ij,j}=0$. By treating $h_{ij}$ as a quantum field, we obtain the equation of motion for freely propagating GWs without the anisotropic stress of the energy-momentum tensor ($\Pi_{ij}=0$). Thus, the equations of motion for the tensor modes $h_k(\tau)$ in Fourier space are described by

\begin{equation}
    \Bigg[\frac{d^2}{d\tau^2}+\frac{2}{a}\frac{da}{d\tau}\frac{d}{d\tau}+k^2\Bigg]h_k(\tau)=0,
\end{equation}
where $k$ is the wave number, and $\tau$ represents the conformal time with its derivative $d\tau=dt/a$. The current energy spectrum of primordial GWs is given by the present-day tensor power spectrum $\Delta_h^2(k,\tau_0)$

\begin{equation}
    \Omega^{\rm{GW}}_0(f)=\frac{1}{12} \left(\frac{2 \pi f}{H_0}\right)^2 \Delta_h^2(k,\tau_0), \quad f=\frac{k}{2 \pi a_0}.
    \label{gw}
\end{equation}

The wave numbers $k$ are outside the horizon during inflation, implying that the tensor modes $h_k(\tau)$ are independent of the conformal time $\tau$ until the wave numbers $k$ re-enter the horizon. Consequently, the present-day tensor power spectrum is related to the primordial tensor power spectrum through the tensor transfer function $T_h(k)$~\cite{Boyle:2007zx}

\begin{equation}
    \Delta_h^2(k,\tau_0)=T_h(k) \Delta_h^2(k,\tau_i).
\end{equation}

The transfer function behaves as $T_h(k) \sim (a_k^2/a_0^2)$~\cite{Boyle:2005se,Turner:1993vb,Watanabe:2006qe}, where $a_k=k/H(a_k)$ is the scale factor at the $k$-mode horizon re-entry and depends on the equation of state (EOS) at the time of horizon re-entry. More details about the tensor transfer function can be found in Refs.~\cite{Boyle:2007zx}. The calculations of the primordial tensor power spectrum are conventionally based on the CMB pivot wave number $k_{\rm{cmb}}$. Eventually, the current energy spectrum $\Omega^{\rm{GW}}_0(f)$ is described by the primordial tensor amplitude $A_t$,
\begin{equation}
    \Omega^{\rm{GW}}_0(f)=\frac{1}{12} T_h(k) \left(\frac{2 \pi f}{H_0}\right)^2 A_t\left(\frac{k}{k_{\rm{cmb}}}\right)^{n_t},
\end{equation}
where $n_t$ is the tensor spectral index. The primordial GW signals are strongly enhanced by both primordial tensor amplitude $A_t$ and tensor spectral index $n_t$. We therefore reasonably set the maximum $A_t$ and $n_t$ so that the present-day and future GW detectors can detect the primordial GW signals as much as possible. The maximum $A_t$ depends on the amplitude of the primordial scalar spectrum $A_s$ and the tensor-to-scalar ratio $r=A_t/A_s \lesssim 0.07$~\cite{Planck:2018jri}. According to the measured primordial scalar spectrum $A_s=2.1 \times 10^{-9}$, the maximally allowed $A_t$ is set to $1.5 \times 10^{-10}$. On the other hand, we take the similar maximum $n_t=0.4$ as discussed in Refs~\cite{DEramo:2019tit}. Besides, the primordial GW power spectrum decreases with the increase of $r$. Therefore, we investigate the primordial GW power spectrum without an EMD era by fixing $r=0.07$ for detectability. We find that the GW primordial spectrum generated by $n_t<0.13$ is beyond the range of the cosmic explorer(CE)~\cite{LIGOScientific:2016wof} detector, while that produced by $n_t<-0.1$ is beyond the detection capability of the big bang observer(BBO)~\cite{Crowder:2005nr,Corbin:2005ny,Yagi:2011wg} detector.

The EMD era leaves distinct imprints on the primordial GW spectrum, where two characteristic features can be identified - the EMD bump and the matter-dominated tail:
\begin{itemize}
    \item EMD Bump: During the EMD era, the background expansion of the Universe transitions from an accelerated phase to a decelerated phase. This changes the tensor transfer function from $T_h(k) \propto f^{-2}$ to $T_h(k) \propto f^{-4}$, resulting in one kink and a temporary enhancement in the primordial GW spectrum at certain frequencies. This feature is known as the EMD bump. The EMD bump arises due to the presence of additional sources of anisotropic stress during the EMD era, such as collisionless particles or cosmic defects. The amplitude and shape of the EMD bump depend on the properties of these additional sources. The EMD bump is typically characterized by its peak frequency $f_{\rm{peak}}$ and its amplitude relative to the inflationary background.
    \item Matter-Dominated Tail: After the EMD era, the Universe enters the regular radiation-dominated era, during which the primordial GW spectrum continues to evolve. The tensor transfer function evolves from $T_h(k) \propto f^{-4}$ to $T_h(k) \propto f^{-2}$, leading to another kink in the primordial GW spectrum at certain frequency. Also, the amplitude of the primordial GW spectrum decreases as $\Omega^{\rm{GW}}(f) \propto f^{n_t}$ for frequencies below the peak frequency $f_{\rm{peak}}$. This power-law behavior is known as the matter-dominated tail of the primordial GW spectrum. The matter-dominated tail extends to lower frequencies and provides a unique signature of the early universe dynamics.
\end{itemize}
The detection and characterization of the EMD bump and the matter-dominated tail in the primordial GW spectrum can provide valuable insights into the physics of the early universe, including the nature of the EMD era and the generation mechanisms of primordial GWs.
The EMD bump and matter-dominated tail occur
at these two characteristic temperatures $T_{\rm eq}$ and $T_R$ respectively, which single out two corresponding frequencies $f(T_{\rm eq})$ and $f(T_R)$.
The frequency $f(T_R)$ at temperature $T_R$ can be obtained by the Eq.~\ref{gw},
\begin{eqnarray}
    f(T_R)&=&\frac{a(T_R)H(T_R)}{a_0 H_0} f(T_0)  \\
    &=&\left(\frac{g_s(T_0)}{g_s(T_R)}\right)^{1/3}\left(\frac{g_{\rho}(T_R)}{g_{\rho}(T_0)}\right)^{1/2} \left(2\Omega_R^0\right)^{1/2} \frac{T_R}{T_0} f(T_0), \nonumber
    \label{Tfrelation}
\end{eqnarray}
where $T_0=2.72548$ K~\cite{Fixsen:2009ug} is the present temperature of cosmic microwave background(CMB)
and $f(T_0)=H_0/2 \pi$ denotes the current frequency of the GW mode. $H_0=100 h$ km/s/Mpc is the present-day Hubble expansion rate, and $\Omega_R^0$ represents the ratio of the current radiation energy density and critical density. Likewise, we can derive the frequency $f(T_{\rm{eq}})$ by exploiting the relationship between temperature and frequency
\begin{equation}
    f(T_{\rm{eq}})=\left(\frac{g_{\rho}(T_R)}{g_{\rho}(T_{\rm{eq}})}\right)^{-1/6}\left(\frac{T_{\rm{eq}}}{T_{R}}\right)^{2/3} f(T_R).
\end{equation}

\begin{figure}
\centering
\includegraphics[width=1\linewidth]{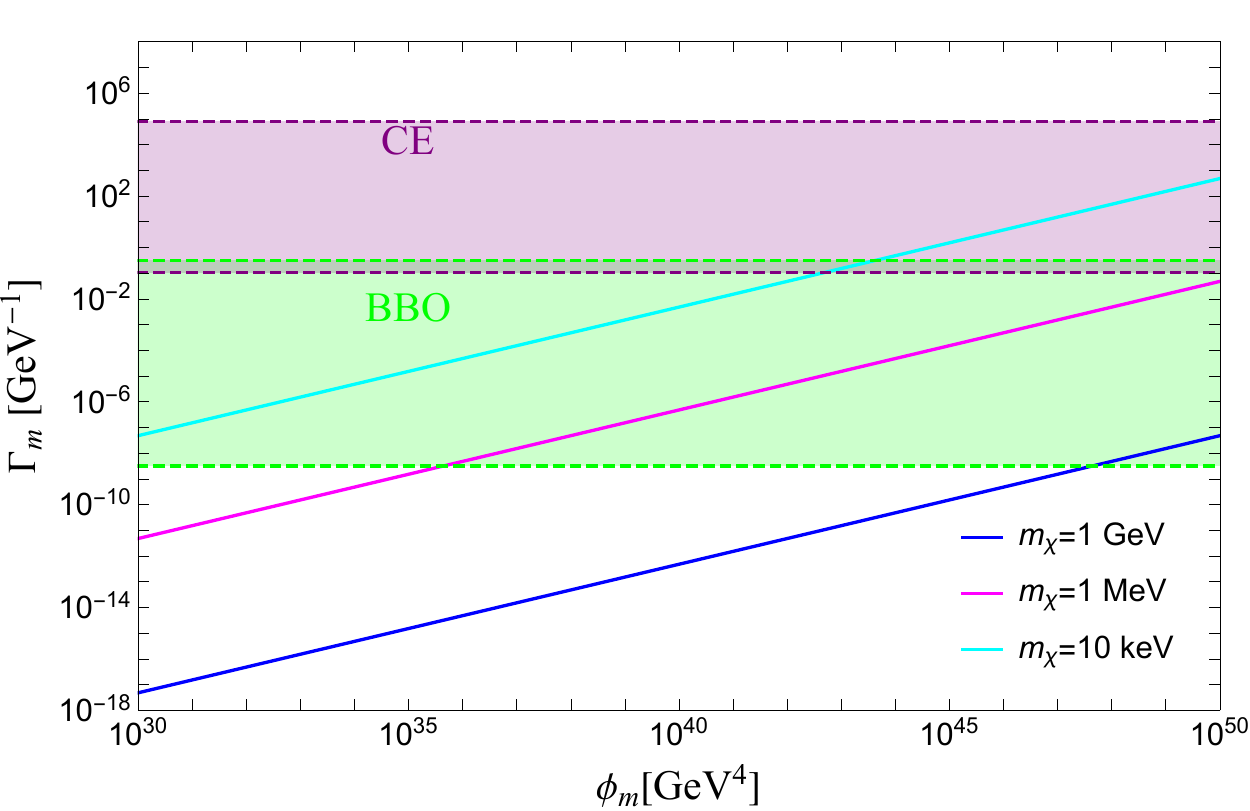}
\caption{The contour plots of observed DM relic density with different masses $m_\chi$ as a function of decay width $\Gamma_m$ and the co-moving energy density $\phi_m$. The blue, magenta, and cyan contour plots are for $m_\chi=10$ keV, $m_\chi=1$ MeV and $m_\chi=1$ GeV respectively. The purple (green) shaded region indicates the case where the lower kink frequency $f(T_R)$ is within the sensitivity of the CE (BBO) detector.}
\label{Lgammaphim}
\end{figure}
\begin{figure}
\includegraphics[width=0.95\columnwidth]{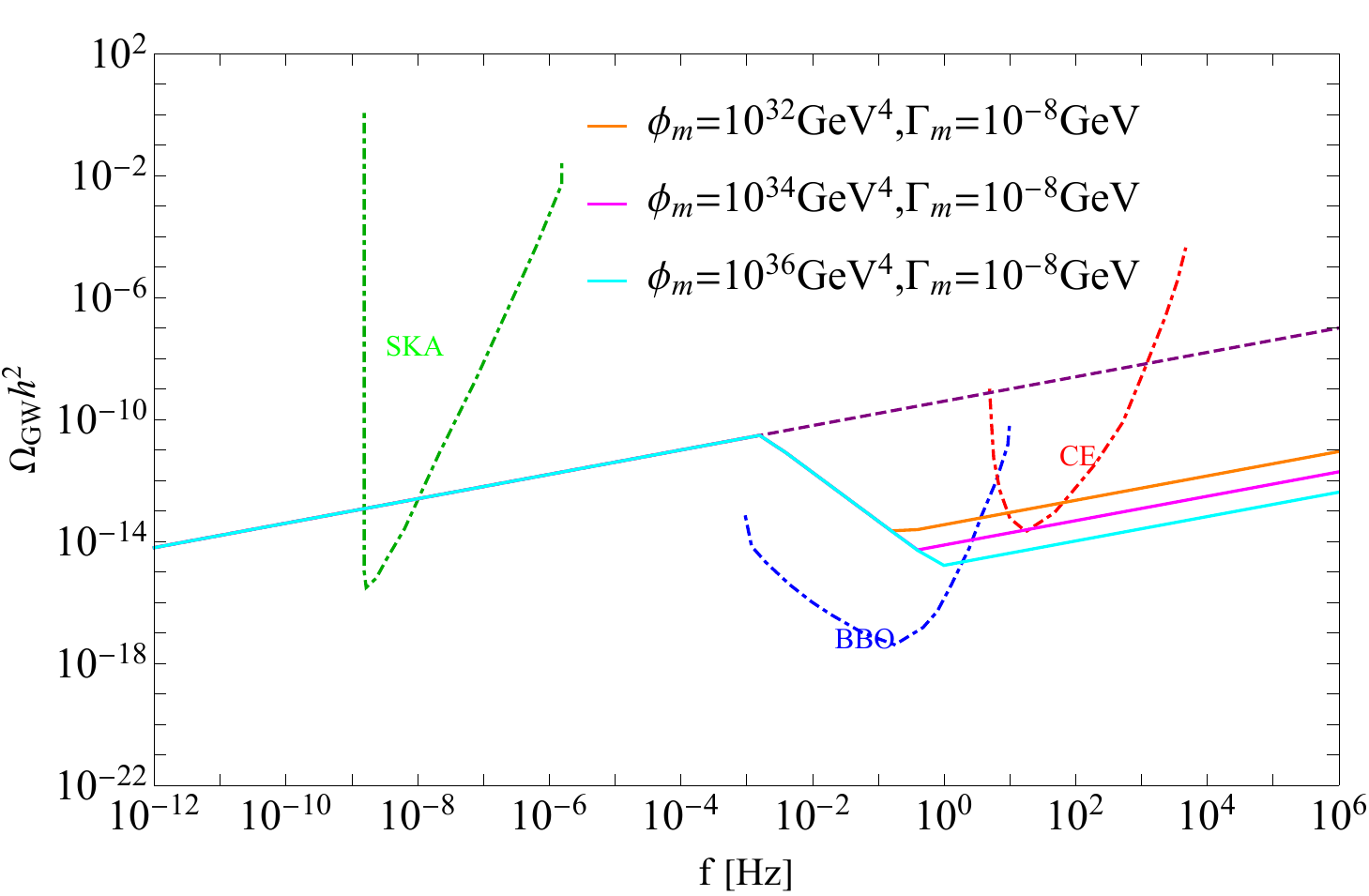}
\includegraphics[width=0.95\columnwidth]{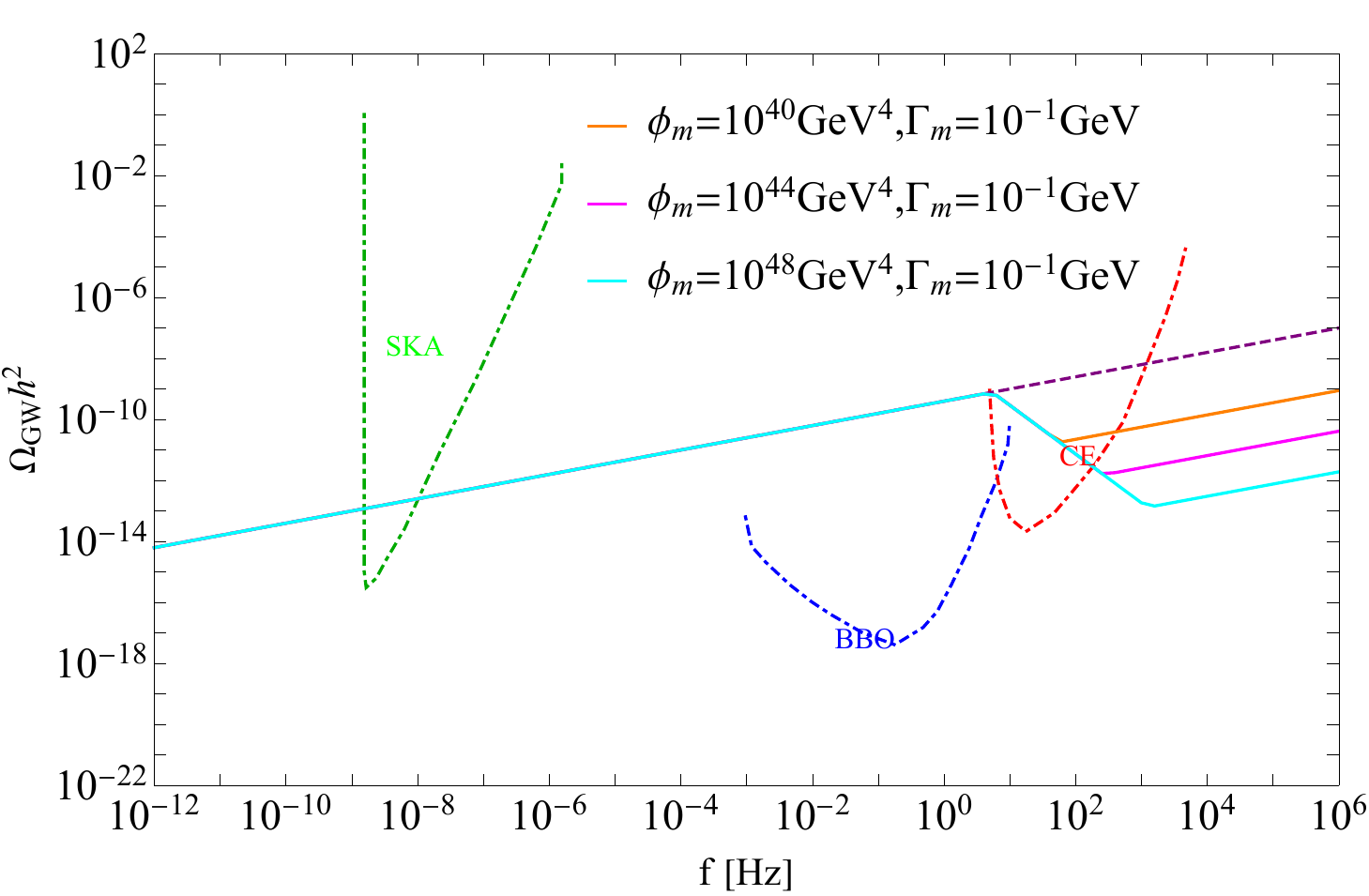}
\caption{The present-day primordial GWs spectra $\Omega_{GW} h^2$ as a function of the frequency. In both the upper and bottom panels, the solid lines illustrate the primordial GW spectra within the EMD era induced by different relevant parameters, while the purple dashed line represents that without the EMD era. The dot-dashed lines show the sensitivity of various GW detectors, such as SKA~\cite{Carilli:2004nx,Janssen:2014dka,Weltman:2018zrl}, BBO~\cite{Crowder:2005nr,Corbin:2005ny,Yagi:2011wg}, and CE~\cite{LIGOScientific:2016wof}. The upper(bottom) panel shows that the primordial GW spectra lie in the sensitivity of BBO(CE). }
\label{LBBOCE}
\end{figure}
In Eq.~\ref{Teq} and Eq.~\ref{TR}, these two characteristic temperatures $T_R$ and $T_{\rm{eq}}$ are only functions of decay width $\Gamma_m$ and the co-moving energy density $\phi_m$ respectively,
which ensures that the frequency $f(T_R)$ is function of $\Gamma_m$ only while $f(T_{\rm eq})$ is function of both $\Gamma_m$ and $\phi_m$.
Hence, we first study the case where the lower frequency $f(T_R)$ is detected by GW detectors such as the CE and BBO as it is only connected with $\Gamma_m$.
In Fig.~\ref{Lgammaphim}, we show, by the shaded region, the parameter space that leads to the lower kink frequency $f(T_R)$ that can be covered by BBO or CE GW experiments. The parameter regions that can provide observed DM relic density for different choices of DM mass are also indicated by the solid lines with different colors.
More importantly, the behavior of transfer function $T_h(k)$ is modified by the EMD era, which eventually leads to a change in the spectral index of GW spectrum
\begin{equation}
    n=n_t+ 2\times\frac{3\omega-1}{3\omega+1},
\end{equation}
where $\omega=0$ arises from the EOS during an EMD era~\footnote{It should be noted that the non-standard cosmological models with EOS $\omega=0$ will leave similar signatures on the gravitational spectrum compared to the EMD era. This indicates that the EMD era may be difficult to distinguish from the phase transition or more complex inflationary dynamics in this scenario.}. Therefore, the spectra index of GW spectrum changes by $\delta n=-2$ during an EMD era. This implies the GW spectrum drops sharply within the EMD era and has a step-like feature between the frequencies $f(T_R)$ and $f(T_{\rm{eq}})$. The GW energy density, which is also diluted by the entropy produced during the EMD era, at the frequencies of these two kinks are approximately related to the dilution factor by~\cite{DEramo:2019tit}
\begin{equation}
    \frac{\Omega_{GW}^0(f(T_{\rm{eq}}))}{\Omega_{GW}^0(f(T_R))} \approx \left(\frac{f(T_{\rm{eq}})}{f(T_R)}\right)^{n_t} D^{-4/3}.
\end{equation}
As mentioned before, the thermally produced DM relic density with a larger mass requires a larger dilution factor from the EMD era to reconcile with the observations.
On the other hand, the EMD era also imprints unique features in the GW spectra.
Therefore, probing the light feebly interacting DM and the measurement of the GW spectrum are strongly correlated. In Fig.~\ref{LBBOCE}, we show the primordial GW spectra for different parameters that are within the sensitivity of BBO (blue dashed line) and CE (red dashed line) experiments. The kinks of the primordial GW spectrum imply the onset and end of the EMD era. Furthermore, one can infer the EOS $\omega=p/\rho$ during the early Universe according to the slopes of the kinks.
As the lower kink frequency $f(T_R)$ only relies on the decay width of moduli $\Gamma_m$, the locations of kinks at lower frequencies are the same for the same decay width $\Gamma_m$. However, the higher kink frequency $f(T_{\rm eq})$ and dilution factor $D$ are related to both the co-moving energy density $\phi_m$ and decay width $\Gamma_m$ of moduli. Additionally, they both increase with the co-moving energy density of moduli $\phi_m$ increasing.
Consequently, the GW spectrum has strong dilution behavior, which is correlated with the light feebly interacting DM mass through the observed relic density.
Provided that the GW spectrum can be exactly observed by GW detectors, the light feebly interacting DM mass $m_\chi$ can also be accurately predicted in this scenario. Besides, the light feebly interacting DM and standard model particles in the general EMD era models can be simultaneously produced by the long-lived parent particle decay at collider. The long-lived parent particle decay leads to various displaced collider signals such as displaced vertices, displaced jets/leptons and stopped particle decays. Therefore, one can also infer the DM mass from displaced signatures at collider\cite{Co:2015pka}.

{\bf{Benchmark Model: Supersymmetry DFSZ axion model}}

\begin{figure}
\includegraphics[width=0.95\columnwidth]{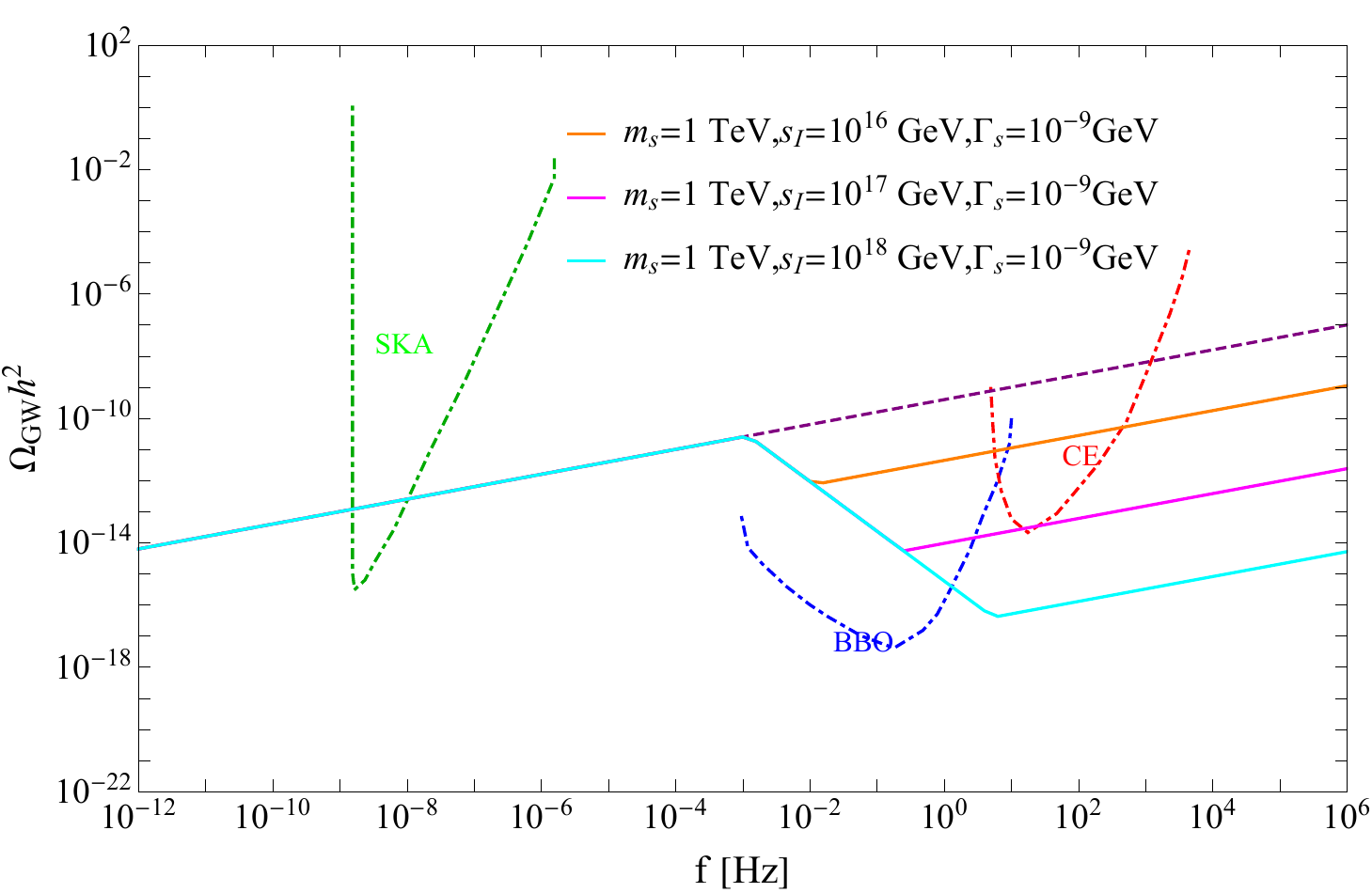}
\includegraphics[width=0.95\columnwidth]{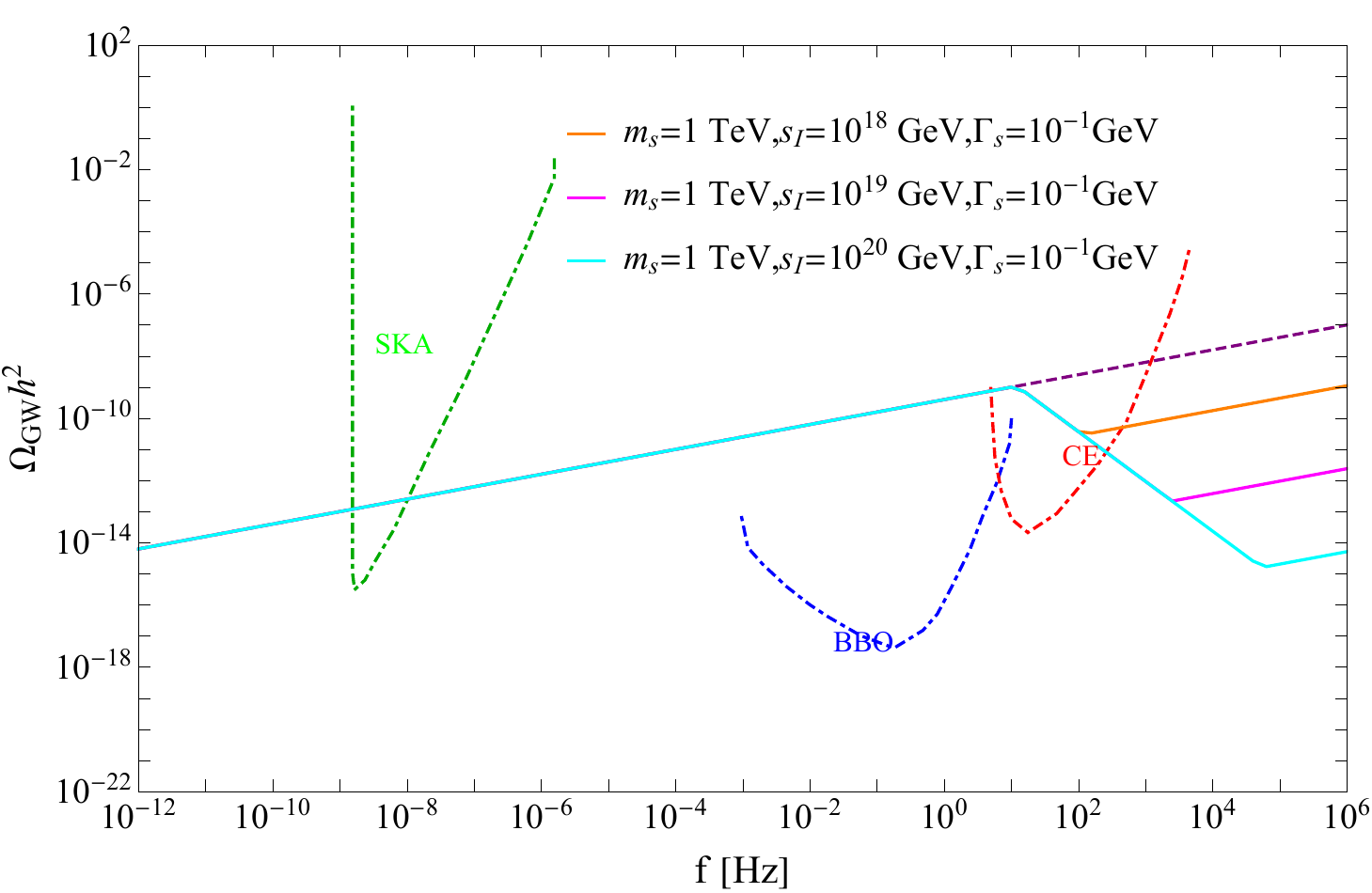}
\caption{The present-day primordial GWs spectra $\Omega_{GW} h^2$ in the SUSY DFSZ axion model as a function of the frequency. These parameters related to saxion are shown. In both the upper and bottom panels, the solid lines illustrate the primordial GW spectra within the EMD era induced by different relevant parameters, while the purple dashed line represents that without the EMD era.}
\label{LBBOCEs}
\end{figure}
In the SUSY DFSZ axion model, axino, the fermionic superpartner of the axion, is a natural light feebly interacting dark matter candidate. The axino would freeze out at the cosmic temperature
\begin{equation}
    T_{\tilde{a}}^f \approx 10^{10} {\rm{GeV}} \left(\frac{V_{\rm{PQ}}}{10^{12}{\rm{GeV}}}\right)^2 \left(\frac{1}{\alpha_s}\right)^3,
\end{equation}
where $V_{\rm{PQ}}$ is the PQ symmetry breaking scale and $\alpha_s$ is the strong interaction coupling constant. When the reheating temperature $T_{R}^{I}$ after inflation is larger than axino freeze-out temperature $T_{\Tilde{a}}^f$, the axino will be in thermal equilibrium and its thermal produced relic abundance is
\begin{equation}
    \Omega_{\chi} h^2 = \frac{m_{\Tilde{a}} Y_{\tilde{a}} s_0}{\rho_c} h^2,
\end{equation}
with the axino yield
\begin{equation}
    Y_{\tilde{a}}=\frac{135 \zeta(3) g_{\Tilde{a}}}{8\pi^4 g_s(T_{\tilde{a}}^f)},
\end{equation}
where $g_{\Tilde{a}}=2$ is the internal d.o.f of axino. The axino abundance produced by thermal freeze-out will easily overclose the Universe. Fortunately, there exists the saxion (the real part of the axion field) in the SUSY DFSZ model. The supersymmetry breaking provides the saxion field with a large potential $\rho_s= m_s^2 s_I^2$ after inflation, which leads to the EMD era in the Universe. The onset and end of the EMD era induced by saxion correspond to the cosmological temperatures $T_M$ and $T_{Rs}$,
\begin{equation}
    T_{M}=3\left(\frac{10}{g_{\rho}(T_{M}) \pi^2}\right)^{1/4} \frac{m_{s}^{1/2} s_{I}^2}{M_{\rm{Pl}}^{3/2}},
\end{equation}
\begin{equation}
    T_{Rs}=\left(\frac{90}{\pi^2 g_{\rho}(T_{Rs})}\right)^{1/4} \sqrt{\Gamma_{s} M_{\rm{Pl}}}. 
\end{equation}
where $m_s$ and $\Gamma_s$ are the mass and decay width of the saxion, respectively. The EMD era will result in two kinks on the primordial GWs spectrum, corresponding to the characteristic frequency $f(T_M)$ and $f(T_{Rs})$
\begin{equation}
    f(T_{Rs})=\left(\frac{g_s(T_0)}{g_s(T_{Rs})}\right)^{1/3}\left(\frac{g_{\rho}(T_{Rs})}{g_{\rho}(T_0)}\right)^{1/2} \left(2\Omega_R^0\right)^{1/2} \frac{T_{Rs}}{T_0} f(T_0),
\end{equation}
\begin{equation}
    f(T_{M})=\left(\frac{g_{\rho}(T_{Rs})}{g_{\rho}(T_M}\right)^{-1/6}\left(\frac{T_M}{T_{Rs}}\right)^{2/3} f(T_{Rs}).
\end{equation}
After the end of the EMD era, an amount of entropy is injected into the Universe and dilutes the overproduced axino relic. The dilution effect can be described by the dilution factor,
\begin{equation}
    D_{s}=\left(\frac{9 g_{\rho}(T_{Rs})}{g_{\rho}(T_{M})}\right)^{1/4} \frac{m_{s}^{1/2} s_{I}^{2}}{\sqrt{\Gamma_{s}} 
        M_{\rm{Pl}}^{2}}.
\end{equation}
With these in hand, we can calculate the primordial GW spectrum in the SUSY DFSZ axion model as shown in Fig.~\ref{LBBOCEs}. If these two kinks on the primordial GW spectrum are fully detected by BBO or CE detectors,  we can infer the parameters related to saxion according to the primordial GW spectrum. For example, if the magenta line is observed by BBO detector, we can evaluate the $\Gamma_s=10^{-9}$ GeV and $s_I=10^{17}$ GeV for fixed the saxion mass $m_s=1$ TeV. Under these parameters, the dilution factor $D \approx 1360$ and the axino mass $m_{\Tilde{a}} \approx 326$ keV is determined by the dilution factor according to the observed DM abundance.

\section{Conclusions}
\label{section4}
The exploration of light feebly interacting DM is challenging through traditional DM direct detection experiments due to its feeble interactions. However, gravitational waves (GWs) offer a novel and complementary avenue for probing the early Universe and searching for light feebly interacting DM.
This approach can be utilized in conjunction with collider experiments and indirect/direct detections.
In particular, the entropy production period plays a crucial role in addressing the overabundance of thermally produced light feebly interacting DM relic density.
On the other hand, such an entropy production period also leaves imprints on the primordial GW spectrum, giving rise to distinct features characterized by two kinks with the frequencies corresponding to the starting ($T_{\rm eq}$) and ending ($T_R$) temperature of the entropy production period. This is illustrated in our work by an EMD era induced by the moduli.

The detection and characterization of these modified GW spectra within the sensitivities of GW detectors such as BBO and CE provide a unique opportunity to probe the existence of light feebly interacting DM.
The DM masses $m_{\chi}$ can be further inferred under the same set of parameters.
Moreover, with the continuous improvement in precision, range, and sensitivity of GW detectors, as well as the synergistic combination of different GW experiments, the primordial GW spectrum can be measured more accurately and extensively.
Consequently, probing light feebly interacting DM through the GW spectrum becomes a highly promising avenue in the foreseeable future.

\section{acknowledgments}

LW is supported by the National Natural Science Foundation of China (NNSFC) under grants No. 12335005, No. 12275134 and No. 12147228. 
BZ is supported by NNSFC under grants No. 12275232 and No. 11835005. 
YW is supported by NNSFC under grant No. 12305112 and also would like to thank Nanjing Normal University for its support.

\bibliography{refs}
\end{document}